\documentclass[aps, twocolumn,prx, showpacs, longbibliography, preprintnumbers,amsmath,amssymb,superscriptaddress]{revtex4-1}
\usepackage{amsmath,amsfonts,amssymb,graphics,graphicx,epsfig,color,times,indentfirst,layout,hyperref,subfigure,multirow,bm}
\usepackage{hyperref, soul}
\usepackage{ulem}
\hypersetup{linktocpage,,colorlinks=true}
\usepackage{threeparttable}
\usepackage{dcolumn}
\usepackage{bm}
\usepackage{color}

\begin{document}
    \title{Digital quantum simulation of dynamical topological invariants on near-term quantum computers}
    \author{Huai-Chun Chang}
    \affiliation{Department of Mathematical Sciences, National Chengchi University, Taipei 11605, Taiwan}
    \author{Hsiu-Chuan Hsu}\email{hcjhsu@nccu.edu.tw}
		
    \affiliation{Graduate Institute of Applied Physics, National Chengchi University, Taipei 11605, Taiwan}
    \affiliation{Department of Computer Science, National Chengchi University, Taipei 11605, Taiwan}
    
    \date{\today}
\begin{abstract}
	Programmable quantum processors are suitable platforms for simulating quantum systems, of which topological phases are of particular interest. We simulate the quench dynamics of a one-dimensional system on IBM Q devices. The topological properties of the dynamics are described by the dynamical topological invariants, the dynamical winding number and the time-dependent Berry phase, which are simulated with the quantum circuit model. The results show that despite the noise present in the current quantum computers, the dynamical topological invariants are robust. Moreover, to investigate the influence of open quantum system, we analytically solve the master equation in Lindblad form and show that the dynamical winding number and the change in Berry phase are not affected by the dissipation. This study sheds light on the robustness of topological phases on the noisy intermediate-scale quantum computers. 

\end{abstract}
\maketitle

\section{Introduction}
The most appropriate tool for simulating quantum systems is the quantum computer because the fundamental of nature is quantum mechanics, as proposed by Feynmann in 1982 \cite{Feynman1982}. 
Since then, the scientific and engineering communities have been pursuing the realization of the device and exploring the applications.  
It has been proved that simulations of quantum systems is plausible due to the universality of the gate-based quantum comuting processors \cite{Lloyd1996, Tacchino2019}. Recently, the quantum technologies have experienced a great breakthrough and improved the fabrication of the quantum processors \cite{Arute2019,Jurcevic2021}. 
However, the current quantum computers are at the stage of Noisy Intermediate-Scale Quantum (NISQ) era \cite{Preskill2018}. The quality and number of qubits are still limited.
The capabilities and applications of NISQ computers have been explored in many subjects. For example, several studies have demonstrated the quantum simulation of non-equillibirum quantum systems \cite{Smith2019,Babukhin2020,Fauseweh2021}, many-body states \cite{Smith2019, Rahmani2020, Kirmani2021}, open quantum systems \cite{Head2019,Del2020,Garcia2020,Head2021, Kamakari2021}, dynamical quantum phase transitions \cite{Guo2019} and topological phases of matter \cite{Murta2020, Ji2020,Mei2020,Chen2021,Xiao2021}. 

Moreover, phase factor is one of the most important aspect that distinguishes quantum mechanics from classical mechanics \cite{Dirac1972,Yang2013}. In classical numerical simulations of quantum systems, one could simulate the wave functions. However, experiments can only obtain the squared modulus of the wave function without the information of phase. For a more realistic simulation of quantum systems,  quantum computer is a suitable platform for controlling the quantum states and allowing the measurement of the phase with a careful design of the quantum circuits \cite{Murta2020,Cian2021,Xiao2021}.

The study of the phase factor is the center of the topological matters in modern condensed matter physics. 
The Berry phase, the phase difference of the wave function when the parameters of the system change in a closed path \cite{Berry1984}, has been found to present in many materials. In solids, the Berry phase is the phase difference of the Bloch states as momentum changes in a closed path in the Brillouin zone \cite{Thouless1982a, Kohmoto1985}. 
The Berry phase has direct consequence on the electronic properties of materials \cite{Xiao2010}, such as anomalous Hall effect \cite{Jungwirth2002}, topological insulators \cite{Hasan2010a} and Weyl semimetals \cite{Yan2017}. 

The study of topology has extended to non-equilibrium states. 
Recently, the topological phases of quench dynamics have been explored \cite{Yang2018, Gong2018, Chang2018,Sun2018,Hu2020}. Quench dynamics consider the time evolution of a quantum state, which is the ground state of the initial Hamiltonian $H_i$, under a sudden change of the Hamiltonian to $H_f$. 
As the quantum state evolves unitarily with the quench Hamiltonian $H_f$, the topology of the equilibrium quench Hamiltonian can be detected from the dyanmics, such as the dynamical winding numbers \cite{Zhu2020} and band-inversion surfaces \cite{Zhang2019, Zhang2019b} .
Moreover, by treating time as an additional dimension, the dynamical topological invariants can be defined on a momentum-time torus in quenching one-dimensional systems. 
For example, previous studies have proposed
 time-dependent Berry phase \cite{Hsu2021}, dynamical Chern number \cite{Yang2018, Chang2018,Hsu2021} and the Skyrmion texture in the momentum-time domain \cite{Wang2019b, Guo2019}.

Several studies have studied topological phases on NISQ computers. Viyuela et al. \cite{Viyuela2018} simulated topological phases interacting with thermal baths and observed topological Uhlmann phases. 
Murta et al. \cite{Murta2020} proposed a Berry phase estimation algorithm that removes the dynamical phase while preserves the geometric phase. Mei et al. \cite{Mei2020} showed quantum simulation of topological insulators and the boundary modes.   
 Xiao et al. \cite{Xiao2021} demonstrated the robustness of the topological invariants against noise by computing Chern number on quantum computers. 
  Quantum simulation on NISQ devices opens a new direction for investigating the topological properties of the quantum states. 
   

In this work, we simulate quench dynamics on the cloud quantum computer provieded by IBM Q and found the topological robustness of the topology in quench dynamics on NISQ devices. 
We compare the results of the simulations of quantum circuits on qasm\_simulator provided by the Qiskit API \cite{qiskit} and IBM Q devices. 
The dynamical winding number and Berry phase were computed and shown to reflect the topological properties. Moreover, we consider the dynamics in the open quantum system described by the master equation and show that the topological invariants are robust against dissipation. The analytical results agree with that of quantum simulation. 
The rest of the paper is organized as follows. In sec. \ref{sec:method}, the quench protocol and the quantum circuit model for computing topological properties are introduced. In sec. \ref{sec:result}, the simulation results and analysis are presented. In sec. \ref{sec:master}, the solution and discussion of the master equation for the dynamics in the open quantum system are given. Finally, the conclusion is given in sec. \ref{sec:concl}

\section{The Quench dynamics}\label{sec:method}
In this study, we consider the quench dynamics in one-dimension. The initial states are the ground state of $H_i=-\sigma_z$. Thus, the pseudospins are polarized along the computational basis. The initial quantum state is suddenly quenched by 
the single-particle Su-Schrieffer-Heeger (SSH) Hamiltonian \cite{Su1968} $\mathcal{H}_0(k)$ 
\begin{eqnarray}
\mathcal{H}_0(k)&=&h_x(k)\sigma_x+h_y(k)\sigma_y,
\label{eq:ham_k}
\end{eqnarray}
where
\begin{eqnarray}
h_x(k) &=& g_f - \cos( k )\nonumber\\
h_y(k) &=& \sin( k )
\label{eq:ham}
\end{eqnarray}
$\sigma_{x,y}$ are Pauli matrices and act on the sublattices
$a,b$.  The lattice constant is taken to be one. 
The eigen-energy
is $E_{\pm}=\pm E_k=\pm\sqrt{h_x^2+h_y^2}, =\pm\sqrt{\sin^2(k)+(g_f-\cos(k))^2}$. 
The topology of the Hamiltonian is characterized by winding number $\mathit{w}$. The topological phase is in the regime $|g_f| < 1$ with $\mathit{w}=1$. While the trivial phase is in the regime $|g_f| > 1$ with $\mathit{w}=0$. $g_f=1$ is the phase transition point. The winding number is the number of times the pseudospin winds about the origin on the $x-y$ plane as the parameter $k$ changes from $0$ to $2\pi$ \cite{Vanderbilt2018book}.

  Two topological invariants defined to characterize the dynamics are studied. The first is the dynamical winding number. At $t>0$, the pseudospin starts to precess about $(h_x(k),h_y(k))$. For a fixed time $t$, the trajectory of the pseduospin projected on the $x-y$ plane  as $k$ varies from $0$ to $2\pi$ reflects the topology of the quench Hamiltonian. For the topological quench Hamiltonian, this trajectory makes full revolutions about the origin. In contrast, for the trivial quench Hamiltonian, the trajectory does not encircle the origin. 
  The dynamics of the pseudospin is characterized by the dynamical winding number
  \begin{eqnarray}
  w_{dyn}=\frac{1}{2\pi}\int_0^{2\pi}dk\frac{\partial \eta_{yx}}{\partial k},
  \label{eq:wd}
  \end{eqnarray}
  where 
  \begin{eqnarray}
  \eta_{yx}={\rm Im } \log[ 
  {\langle\sigma_{x}\rangle}+i {\langle {\sigma_{y}\rangle}}
  ].
  \label{eq:eta}
  \end{eqnarray}
  In the noiseless situations, for the quench protocol considered here, $\eta_{yx}$ equals to ${\rm Im}\log[ (h_y-ih_x)(\sin(2E_kt)) ]$ with the Planck constant $\hbar$ is set to one. The details can be found in Appendix \ref{app:timeavg}. 
  Because the principal value of the complex log lies in $(-\pi,\pi]$, when the psuedospin makes one full revolution on the $x-y$ plane, $\eta_{yx}$ shows a discontinuous jump between $-\pi$ and $\pi$. 
  For the topological dynamics with $w_{dyn}=1$, $\eta_{yx}$ shows a discontinuous $2\pi$ jump. In contrast, for trivial dyanmics, $\eta_{yx}$ is a smooth function. 
  
  The dynamical winding number can be regarded as the detection of bulk topology of the quench Hamiltonian. Nevertheless, one can treat time as an additional dimension and define a topological invariant on the momentum-time space. 
  This topological invariant is the dynamical Chern number, which is defined on the momentum-time torus which $k\in [0,2\pi], t\in [0,\pi/(2E_k)]$. Since the topology is robust against smooth deformation, it is equivalent to rescaling $E_k=1$, for which the torus becomes $k\in [0,2\pi], t\in [0,\pi/2]$ \cite{Yang2018, Chang2018,Guo2019}.
  The dynamical Chern number can be obtained by integrating the time derivative of the Berry phase \cite{Gresch2017,Kuno2019,Hsu2021}
\begin{eqnarray}
C_{dyn}=\frac{1}{2\pi}\int^{\pi/2}_0 dt \frac{\partial\gamma(t)}{\partial t},
\label{eq:cdyn}
\end{eqnarray}
where $\gamma(t)$ is the time-dependent Berry phase given by the overlap matrix between neighboring k-points \cite{Vanderbilt2018book,Kuno2019}
\begin{eqnarray}
\gamma(t)&=&{\rm Im } \log[\prod_{k}M_{k,k+\delta k}],\nonumber\\
M_{k,k+\delta k}(t)&=&
\langle \psi(k)|\psi(k+\delta k)
\rangle. 
\end{eqnarray} 
$|\psi(k)\rangle$ is the quantum state of the pseudospin at time $t$ with $E_k=1$. 
The Berry phase is the total accumulated phase as the parameter $k$ changes from $0$ to $2\pi$. 
Analytically, the Berry phase is shown to be  $\gamma(t)=-2\pi\mathit{w}\sin^2 t$ (details are given in Appendix \ref{app:berry}), proportional to the winding number of $\mathcal{H}_0$ \cite{Chang2018}. The change in Berry phase from $t=0$ to $t=\pi/2$ is $2\pi\mathit{w}$.
Due to the interval of the principal value of the complex log, the Berry phase $\gamma(t)$ shows a discontinuous jump between $-\pi$ and $\pi$ for topological dynamics, while exhibits a smooth function for trivial dynamics. 
 Thus, we compute the Berry phase and determine the topology by the Berry phase flow (the change of Berry phase in a period taking into account the $2\pi$ jump).

\begin{figure}
	\includegraphics[scale=0.25]{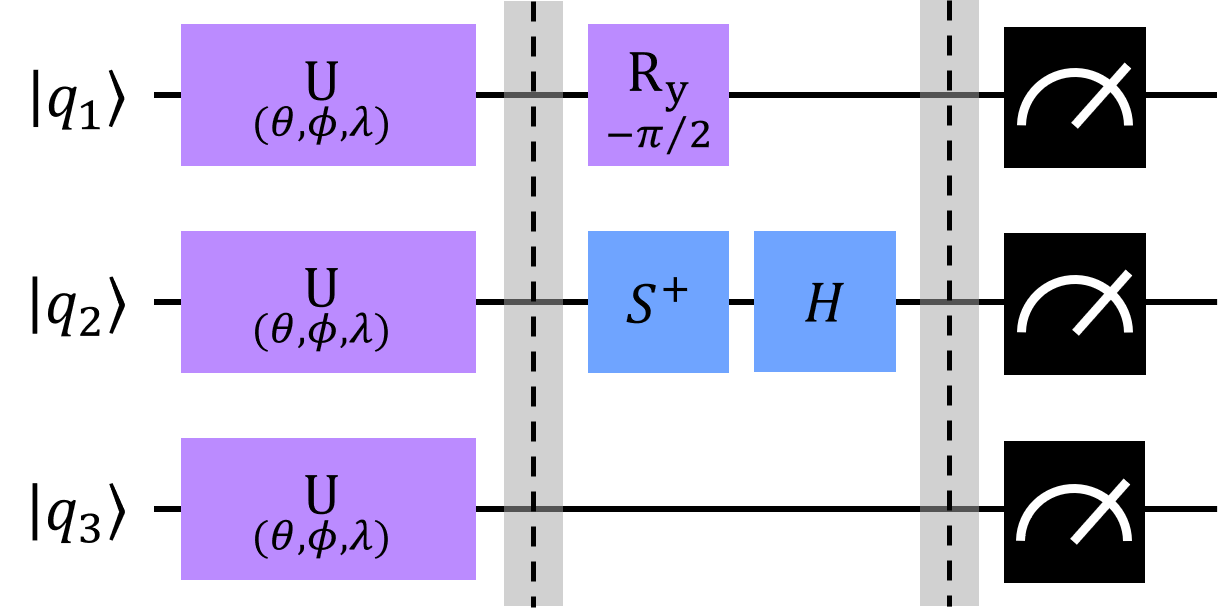}
	\caption{ The circuit model for measuring the time evolution of the pseudospinors. 
		The rotation angles for the unitary gates are $\theta = 2E_kt_i\Delta t$, where $\Delta t = \frac{\pi}{T},\ T=40 $ and $\ t_i \in \{1, 2, 3, \cdots, 20\}$,
		$\phi = - {\rm Im}(\log(h_x+ih_y)) - \frac{\pi}{2}, \lambda=-\phi$.
		$R_y$ is the rotation gate about $y$-axis, $S^{\dagger}$ is the Hermitian conjugate of the phase gate and $H$ is the Hadamard gate.
		The three qubits are used to calculate $\langle\sigma_{x,y,z}\rangle$, respectively. 
	}
	\label{fig:unitary}
\end{figure}
\begin{figure}
	\includegraphics[scale=0.18]{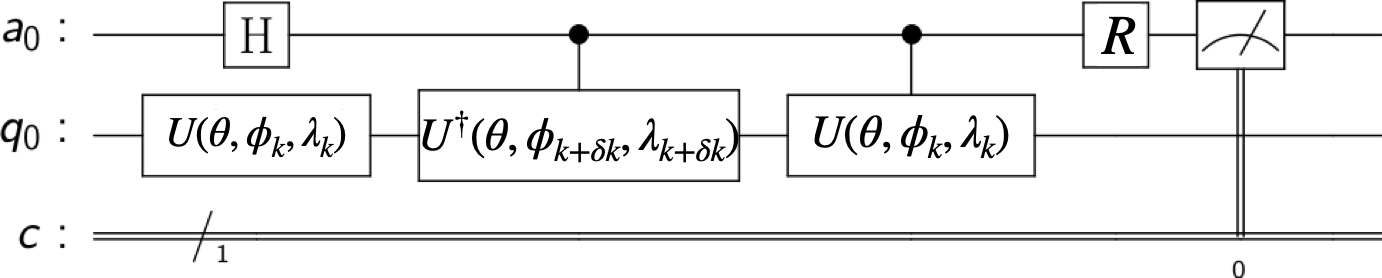}
	\caption{The quantum circuit for calculating the phase between two adjacent momenta. $a_0$ is the ancilla bit for measuring the phase.  $q_0$ is the target bit of which the quantum state represents the quench dynamics. $H$ gate is the Hadamard gate, $R$ gate refers to $+(-)\pi/2$ rotation about $y(x)$-axis when measuring the real(imaginary) part of the phase. For the unitary gates, $\theta=2t, \phi_k = - {\rm Im}(\log(h_x(k)+ih_y(k))) - \frac{\pi}{2}$ and $\lambda_{k}=-\phi_{k}$.}
	\label{fig:berrycirc}
\end{figure}

  The dynamical winding number and Berry phase are computed on the quantum computers and quantum circuit simulators on a classical device. First, we compute the in-plane pseudospins to obtain the dynamical winding number. The quantum circuit for this purpose is shown in Fig. \ref{fig:unitary}. We map the time evolution operator to the U-gate provided by qiskit API. The details can be found in Appendix \ref{app:timeavg}. Because the measurement is in the computational basis, i.e. along $z$-axis, a rotation on the quantum state is performed before the measurement of $\langle \sigma_{x,y}\rangle$. Second, we compute the Berry phase flow. 
  We added an ancilla bit and apply the control unitary bewteen the ancilla and target bit to compute the phase difference between state at $k$ and $k+\delta k$ \cite{Xiao2021}. 
  To aqcuire the real(imaginary) part of the phase, a $(-)\pi/2$ rotation about $y(x)$-axis is performed before the measurement, as shown in Fig.\ref{fig:berrycirc}. It is equivalent to the measurement of the expectation values of $ \sigma_{x,y}$, respectively. 
  From the measurement outcome of the circuit, the Berry phase is obtained by 
  \begin{eqnarray}
  \gamma(t)={\rm Im } \log[ \prod_k {\langle\sigma_{x}\rangle_k}+i {\langle {\sigma_{y}\rangle_k}}], \label{eq:gamma}
  \end{eqnarray}
  where $\langle\sigma_{x,y}\rangle_k=\langle \psi(k)|\psi(k+\delta k)\rangle$ computed with the quantum circuit. 
  In our simulation for Berry phase, $E_k$ is normalized to $1$ for the same reason as mentioned in the previous paragraph.

\section{Results of the Digital quantum simulations}\label{sec:result}
In this section, we present the results of quantum simulation and discuss the quench dynamics of the pseudospins. To investigate whether the Berry phase flow is robust on NISQ computers, we choose the quench Hamiltoians to be near the topological phase transition. In all the experiments, the number of samplings is taken to be 8192, the maximum shots on IBM Q machines. 
\subsection{Dynamical winding number}
\begin{figure}
		\includegraphics[scale=0.17]{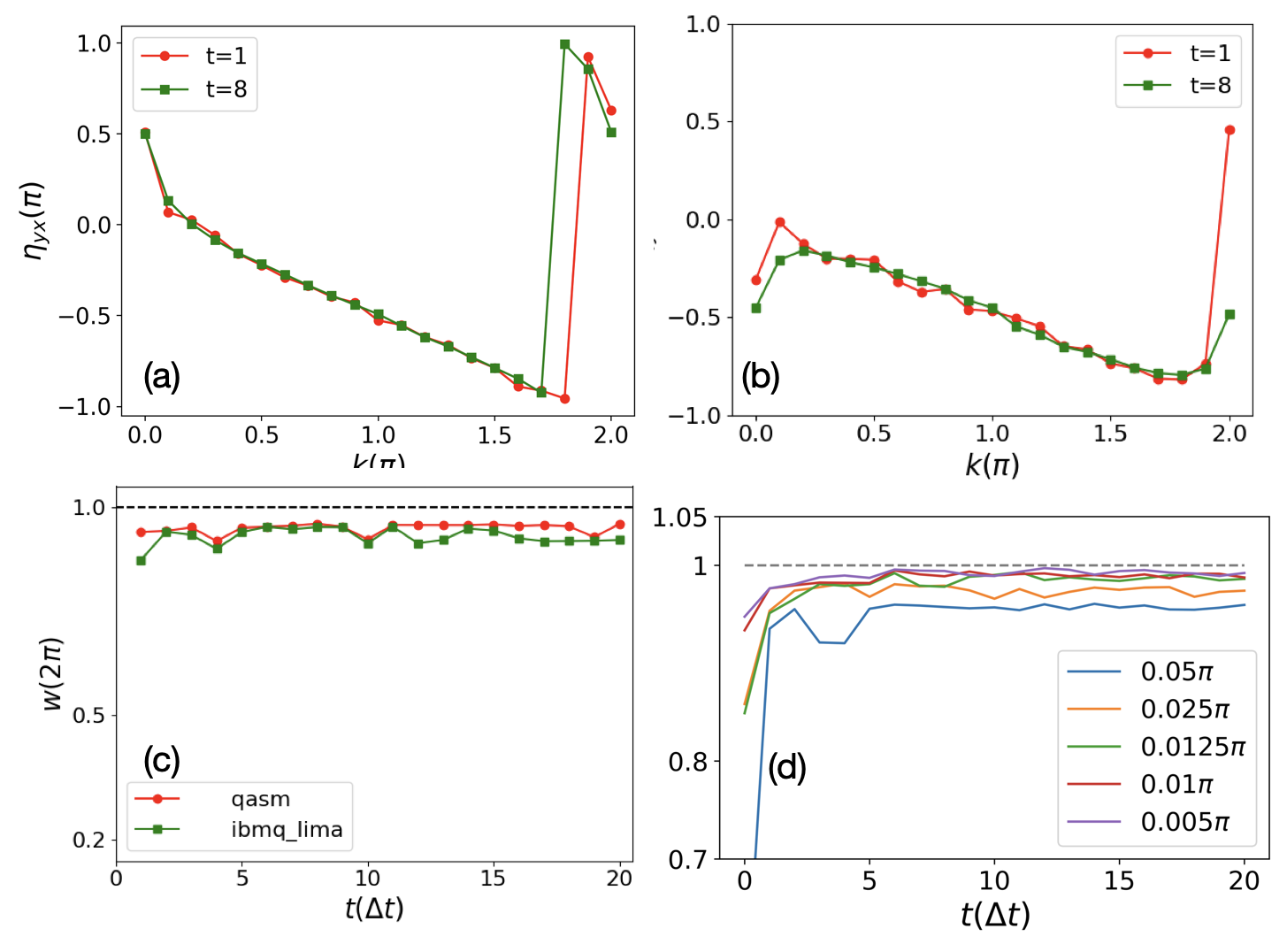}
\caption{(a)The winding angle at $t=\pi/40$ and $t=8\pi/40$ obtaind with ibmq\_lima for quench Hamiltonian $g_f=0.8$. (b) The same as (a) but for $g_f=1.2$.  (c) The winding number extracted from the discontinuity of $\eta_{yx}$ simulated on different devices. (d) The convergence test for the dynamical winding number on qasm simulators. The curves are for different $k$ mesh as shown in the legend.}
\label{fig:dwn}
\end{figure}

Since the winding number corresponds to that the in-plane pseudospins winds around the origin by multiples of $2\pi$, we monitor the dynamical azimuthal angles swept by the pseudospins.  
Fig. \ref{fig:dwn}(a,b) shows $\eta_{yx}$ simulated on ibmq\_lima. In (a), the quench Hamiltonian is topological with $g_f=0.8$.
A discontinuous jump can be observed near $k\approx 1.8\pi$, indicating the full revolution on the $x-y$ plane. 
  In Fig. \ref{fig:dwn}(b), the quench Hamiltonian is trivial with $g_f=1.2$. There is no $2\pi$ discontinuous jump for the trivial case. 
 
 From the dynamical azimuthal angles, we obtain the dynamical winding number. In principle, the dynamical winding number is obtained with Eq. (\ref{eq:wd}) which requires dense mesh in digital simulations. 
 Because of the limited availabitliy of the NISQ devices, we only use sparse mesh in quantum simulation. Thus, we extract the difference of the jump of $\eta_{yx}$ at the discontinuity in unit of $2\pi$ as the dynamical winding number. The results are presented in Fig. \ref{fig:dwn} (c). The results do not show significant deviation between the qasm\_simulator and ibmq\_lima.
 The winding number is not exactly quantized to unity since the $k$ mesh is not dense enough. We performed convergence test on the qasm\_simulator. Fig. \ref{fig:dwn}(d) shows that as the $k$ mesh density increases, the dynamical winding number becomes closer to quantization. Even though the dynamical winding number is not well quantized due to sparse mesh, the dynamical azimuthal angles show the a $\sim2\pi$ jump as a signature of topological dynamics.

\subsection{Berry phase}
\begin{figure}[ht]
	\includegraphics[width=0.5\textwidth]{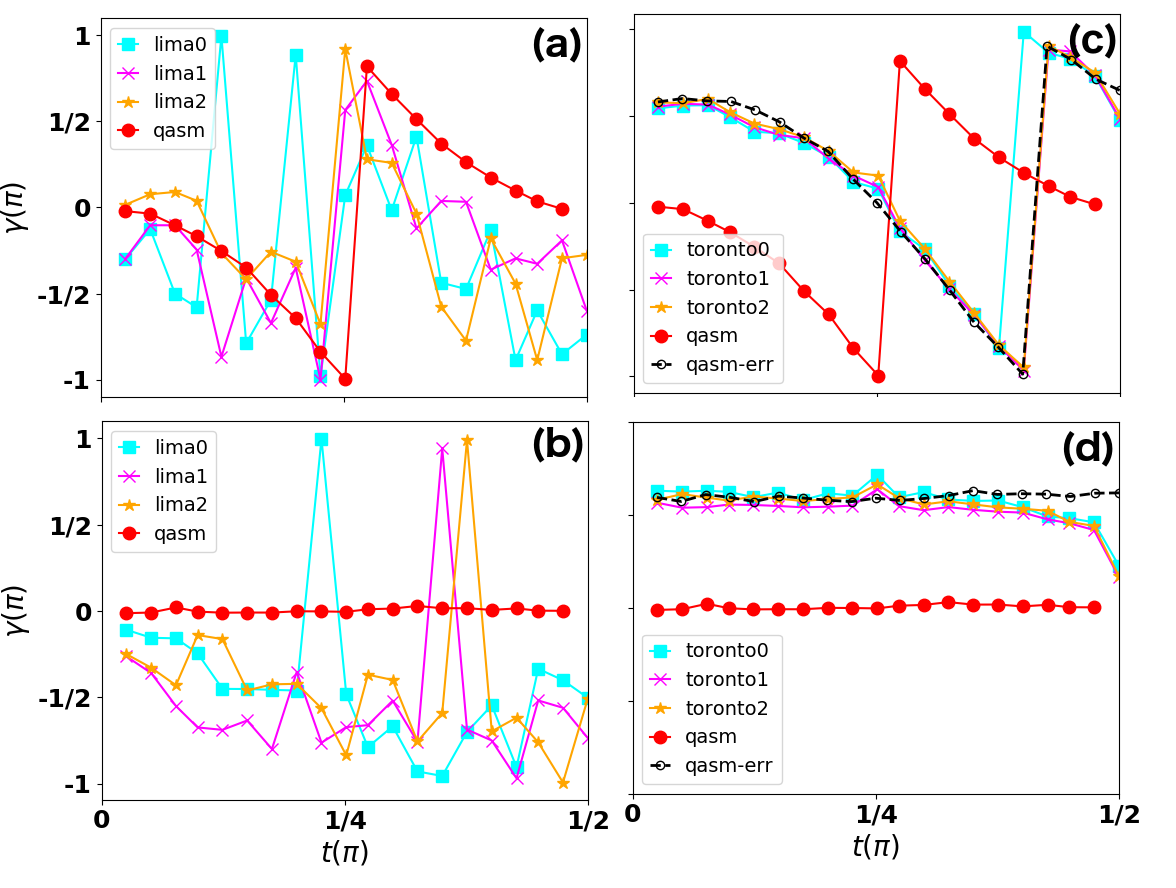}
	\caption{Berry phase for topological quench dynamics.
		(a, b) The Berry phase obatained on ibmq\_lima on for $g_f=0.8, 1.2$, respectively. 
		(c, d) The Berry phase obtained on ibmq\_toronto for $g_f=0.8, 1.2$, respectively. qasm-err denotes the results for simulating the unitary errors on the qasm\_simulator. 
	}
	\label{fig:berrysim}
\end{figure}
We execute the quantum circuit (Fig. \ref{fig:berrycirc}) to compute the time-dependent Berry phase. The circuits are performed on noiseless qasm\_simulators and IBM Q machines.   
The results for $g_f=0.8$ and $g_f=1.2$ are presented in Fig. \ref{fig:berrysim}. 
For qasm\_simulators, as indicated by the red curve with circles,  the $2\pi$ jump is exactly at $t=\pi/4$ for topological dynamics. We present the results on two IBM Q devices, ibmq\_lima and ibmq\_toronto. For each device, we present results of three experiments taken on the same day. Different runs of data show the same behavior. 
For ibmq\_lima, the results for $g_f=0.8$, as presented in Fig. \ref{fig:berrysim}(a),  show a clear discontinuity near $t=\pi/4$, giving rise to a nearly $2\pi$ Berry phase flow from $t=0$ to $t=\pi/2$. The first experiment, as labeled by 'lima0' in the figure, shows random spikes. The similar spikes are observed for the trivial case, as shown in Fig. \ref{fig:berrysim}(b). These sudden spikes do not contribute to Berry phase flow in the period of time $[0,\pi/2]$ because the change of the phases on each side of the spike cancels. Thus, the results in \ref{fig:berrysim}(b) indicate trivial dynamics.  

For ibmq\_toronto, the results for $g_f=0.8$  show a clear discontinuity, as shown in Fig. \ref{fig:berrysim}(c). The position of dicontinuity shifts by $\sim\pi/8$ to the right of $\pi/4$.
The shift of the discontinuity is attributed to the unitary error, which is discussed in detail in the next paragraph. There is a background Berry phase $\pi/2$, which is not shown on qasm\_simulator. The background phase does not change the Berry phase flow. 
 The results for $g_f=1.2$ are shown in Fig. \ref{fig:berrysim}(d). The Berry phase is near $\pi/2$ at all instants. There is no Berry phase flow, indicating trivial dynamics.  For both parameters, there is a $\sim\pi/2$ background Berry phase on ibmq\_toronto, but not on ibmq\_lima. 
 
  We simulate the error on qasm\_simulator by adding a global phase $\delta$ to the target bit in the control-unitary gate (Fig. \ref{fig:berrycirc}). This error gives rise to a background Berry phase, but only shifts the position of the $2\pi$ jump by $\sim\pi/16$. To take into account the shift observed in Fig. \ref{fig:berrysim}(c) , an error $\delta_t$ is added to the polar angles in the unitary and controlled-unitary gates. 
 It is found that with $\delta=0.03\pi$
 and $\delta_t=-0.12\pi$,
 the resultant Berry phase captures the trend on ibmq\_toronto, as shown by the black dashed line in Fig. \ref{fig:berrysim} (c,d). 
 Therefore, we conclude that the background Berry phase is a result of the accumulated global phase due to the systematic unitary gate errors. The shift of the $2\pi$ jump is a consequence of both errors in global phase and polar angles.


 Comparing to the results obtained from ibmq\_lima, ibmq\_toronto shows less fluctuations and no random spikes. Among the quantities for device characetization, including average control-not errors, readout errors, depolarizing time ($T_1$) and decoherence time ($T_2$), the largest difference between the two devices is the time constants. 
The average depolarizing and decoherence time for ibmq\_toronto are $T_1=101\mu s, T_2=122\mu s$, while for ibmq\_lima are $T_1=66\mu s, T_2=99\mu s$. The longer $T_1,T_2$ constants preserve the qubit states and allows more precise measurement outcome. Therefore, the cleaner result on ibmq\_toronto is attributed to the longer time constants.

\section{The master equation}\label{sec:master}
\begin{figure}
	\includegraphics[width=0.28\textheight]
	{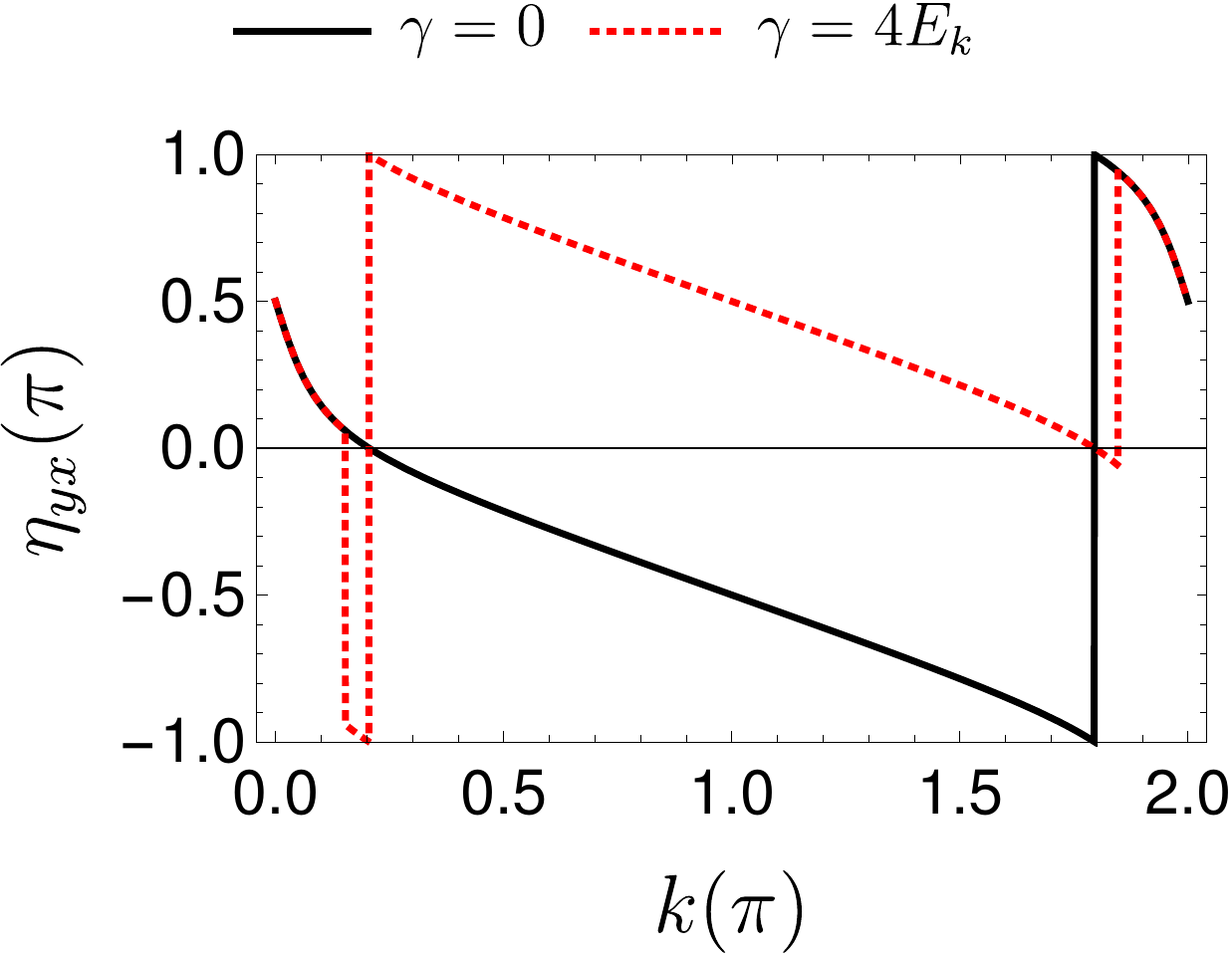}
	\caption{$\eta_{yx}$ as a function of $k$ at $t=0.2\pi$ given by Eq. (\ref{eq:xy}). The overdamping case $\gamma=4E_k$ is shown by red solid line, compared to the dissipationless case with $\gamma=0$ shown by the black solid line.}
	\label{fig:thermal}
\end{figure}

\begin{figure}
	\includegraphics[width=0.37\textheight]{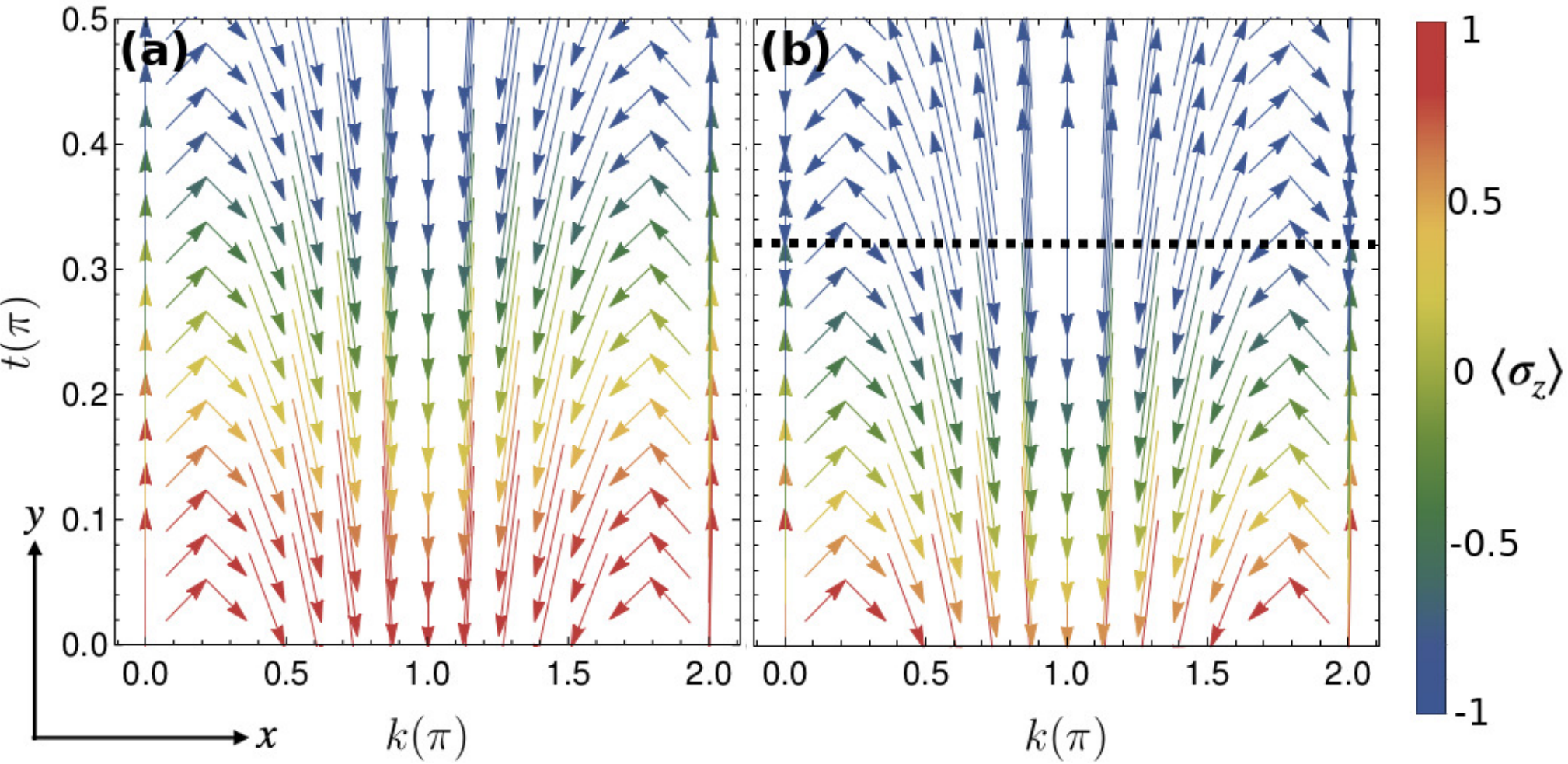}
	\caption{The pseudospin texture in the momentum-time space. (a) Without dissipation, $\gamma=0$. (b) $\gamma=0.5 E_k$. The color denotes the magnitude of the $z$ components. The arrows represent the $x,y$ direction of the pseudospinors. The black dotted line denotes the boundary of time for the momentum-time torus.}
	\label{fig:spin}
\end{figure}

In experiment settings, the qubits are coupled to environments that lead to dissipation. In order to gain insight on how dissipation affects the quench dynamics, we analytically solve the master equation that describes the open quantum system. After quench, the qubit dynamics evolves according to $\mathcal{H}_0+H_{int}$, where $\mathcal{H}_0$ is given in Eq. (\ref{eq:ham_k}) and $H_{int}$ describes the interaction between the qubit and radiation fields. With dipole approximation and rotating wave approximation, the interaction is written as $H_{int}=\vec{d}(e^{i\Omega t}\sigma_++e^{-i\Omega t}\sigma_-)$, where $\vec{d}$ is the tansition matrix element of the dipole operator, $\sigma_{\pm}=(\sigma_x\pm i\sigma_y)/2$ denotes transition between two levels and $\Omega $ is the angular frequency that is the same as the energy gap of the quench Hamiltonian.
With weak coupling limit and Born-Markovian approximations,
the evolution of the density matrix is given by the Lindblad equation \cite{Carollo2003,Breuer}
\begin{eqnarray}
\frac{d\rho}{dt}&=&-i[\mathcal{H}_0(k),\rho]\nonumber\\
&+&\gamma_0(N+1)(\sigma_-\rho\sigma_+-\frac{1}{2}\left\{ \sigma_+\sigma_-,\rho\right\}) \nonumber\\
&+&\gamma_0N(\sigma_+\rho\sigma_--\frac{1}{2}\left\{ \sigma_-\sigma_+,\rho\right\}),
\label{eq:lindblad}
\end{eqnarray}
where $\rho$ is the density matrix, 
$N$ is the Bose-Einstein distribution, $\gamma_0$ is the spontaneous emission rate. 
The Planck constnt $\hbar$ is set to one.  $\gamma_0\sigma_-$ describes the spontaneous emission, $\gamma_0 N\sigma_{-(+)}$ describes the thermally induced emission (absorption). The details of the derivation are shown in Appendix \ref{app:master}. 
After solving for $\rho$ at low temperature, where $\gamma_0\approx \gamma$, the psuedospins are given by
\begin{eqnarray}
\langle \sigma_x(t)\rangle&=&h_y F'(\gamma,k,t)\nonumber\\
\langle \sigma_y(t)\rangle&=&-h_x F'(\gamma,k,t), 
\label{eq:xy}
\end{eqnarray}
where 
\begin{widetext}
	\begin{eqnarray}
	F'(\gamma,k,t)=4e^{-\frac{3\gamma t}{4}}
\frac{(8E_k^2+5\gamma^2)\sin(\frac{\omega t}{4})+\omega\gamma\cos(\frac{\omega t}{4})}
{\omega(4E_k^2+\gamma^2)}
	-\frac{4\gamma}{4E_k^2+\gamma^2}
	\end{eqnarray}
\end{widetext}
and $\omega=\sqrt{64E_k^2-\gamma^2}$. The expression of $F'$ is complicated, but  only its sign affects the measurement of $\eta_{yx}$ [Eq. (\ref{eq:eta})]. As shown in Fig. \ref{fig:thermal}, at a damping rate $\gamma=4E_k$, which is twice the energy gap of the quench Hamiltonian, $\eta_{yx}$ shows a total $2\pi$ difference when $k$ changes from $0$ to $2\pi$.  
To see how dissipation affects the pseudospins' dynamics on the Bloch sphere, we look into the z-component
\begin{widetext}
	\begin{eqnarray}
	\langle\sigma_z\rangle=
	e^{-\frac{3\gamma t}{4}} \frac{2 {\omega} \left(2E_k^2+\gamma^2\right) \cos \left(\frac{{\omega} t}{4}\right)-2 \gamma \left(\gamma^2-14 E_k^2\right) \sin \left(\frac{{\omega} t}{4}\right)}{{\omega} \left(4 E_k^2+\gamma^2\right)}-\frac{\gamma^2}{4E_k^2+\gamma^2}.
	\label{eq:z}
	\end{eqnarray}
\end{widetext}
The geometric phase is a consequence of the dynamics of the pseudospins \cite{Wang2019b}. Thus,
we visualize the pseudospin texture as a function of time and momentum given by Eq. (\ref{eq:xy}) and \ref{eq:z}. As explained in Sec. \ref{sec:method}, $E_k$ is normalized to $1$ in the calculation. Fig. \ref{fig:spin} (a) shows the texture for $g_f=0.8$ in the dissipationless limit when $\gamma$ is set to $0$. At each moment, the $x,y$ components of the pseudospinors winds clockwisely as $k$ varies from $0$ to $2\pi$. While the dissipative case with  $\gamma=0.5E_k$ is shown in Fig. \ref{fig:spin} (b). The $x,y$ components wind clockwisely. The major difference with the dissipationless limit is that the $z$-components decays faster. It shows that the pseudospinor wraps the Bloch sphere for a shorter period of time, as denoted by the black dotted line on the figure. Thus, the $2\pi$ jump in Berry phase can be observed in the presence of dissipation at a shorter time. 


\section{Conclusion}\label{sec:concl}
In this study, we performed quantum simulations of quench dynamics on IBM Q devices.
The dynamical winding number and the Berry phase flow were simulated with quantum circuits and computed on ibmq\_lima and ibmq\_toronto. 
Despite the unitary errors and randomness observed in NISQ processors, the quantum simulation of quench dynamics shows robust signatures for identifying topological and trivial dynamics.  
Moreover, the influence of open quantum system on the topology is addressed in this paper. To study the dissipation, we analytically solve the master equation in the Lindblad form. We found that with dissipation, the dyanmical winding number and Berry phase flow are robust. The analytic solution provides insight on the topological quench dynamics in open quantum systems.

\section*{Acknowledgments}
We would like to thank Po-Yao Chang, Liang-Yan Hsu, Jhih-Shih You and Geng-Ming Hu for valuable discussions. We acknowledge the IBM Q service via NTU-IBM Q hub. This work is supported by the Ministry of Science and Technology (MOST) in Taiwan with Grant No. 108-2112-M-004-002-MY2.





\appendix
\section{Expectation values of pseudospins}\label{app:timeavg}
Given the quench Hamiltonian, the unitary operator for time evolution is 
\begin{eqnarray}
U(t)=\cos(E_kt)-i\sin(E_kt)\hat{h}\cdot
\bm {\sigma }, 
\end{eqnarray}
where $\hat{h}=\hat{h}_x\hat{x}+\hat{h}_y\hat{y}$ with $\hat{h}_{x,y}=h_{x,y}/\sqrt{h_x^2+h_y^2}$.
For simulation on IBM Q simulators and machines, we utilize the U-gate \cite{qiskit}
\begin{eqnarray}
U(\theta,\phi,\lambda)=
\begin{pmatrix}
\cos(\theta/2) & -e^{i\lambda}\sin(\theta/2) \\
e^{i\phi}\sin(\theta/2) & e^{i(\phi+\lambda)}\cos(\theta/2)
\end{pmatrix}
\end{eqnarray}
with $\theta=2E_kt, \phi={\rm Im }\log(h_x + i h_y) - \pi / 2, \lambda=-\phi$. It can be shown that with this choice of parameter, the U-gate is exactly the time evolution operator. 

 For an initial state
\begin{eqnarray}
|\psi_k(t=0)\rangle=
\begin{pmatrix}
a\\
b
\end{pmatrix}
\end{eqnarray}
the time evolution due to the Hailtonian [(Eq. \ref{eq:ham_k})] is given by 
\begin{eqnarray}
	|\psi_k(t)\rangle=
	\begin{pmatrix}
	a\cos(E_kt)-ib\sin(E_kt)(\hat{h}_x-i\hat{h}_y) \\
	b\cos(E_kt)-ia\sin(E_kt)(\hat{h}_x+i\hat{h}_y)
	\end{pmatrix}.
	\label{eq:psi}
\end{eqnarray}

For the initial states consiedered in the main text $(1 \ \ 0)^T$ , the expectation values of spinors are 
\begin{eqnarray}
\langle \sigma_x(t)\rangle&=&\hat{h}_y\sin(2E_kt)\\
\langle \sigma_y(t)\rangle&=&-\hat{h}_x\sin(2E_kt)\\
\langle \sigma_z(t)\rangle&=&\cos(2E_kt).
\end{eqnarray}

Thus, the azimuthal angle swept by the in-plane components is given by ${\rm Im } \log[\langle \sigma_x\rangle+i\langle \sigma_y\rangle]=\log[(h_y-ih_x)\sin(2E_kt)]$.
\section{The analytical expression for the Berry phase}\label{app:berry}
The Berry phase can be calculated analytically by
\begin{eqnarray}
\gamma(t)&=&-{\rm Im }\int_{0}^{2\pi}dk\langle \psi_k(t)|\partial_k\psi_k(t)\rangle.
\end{eqnarray}
It is the continuum version of Eq. (\ref{eq:gamma}) in the main text. 
Using Eq. (\ref{eq:psi}) with $(a\ \ b)=(1 \ \ 0)$ and $E_k=1$, one obtains 
\begin{eqnarray}
\gamma(t)&=&-{\rm Im }\int_{0}^{2\pi}dk\langle \psi_k(t)|\frac{d\phi}{dk}\partial_{\phi}\psi_k(t)\rangle\\
&=&-{\rm Im}\int_0^{2\pi}dk\frac{d\phi}{dk}\sin^2t\nonumber\\
&=&-2\pi\mathit{w}\sin^2 t,
\end{eqnarray}
where $\mathit{w}$ is the winding number of the quench Hamiltonian. Thus, the dynamical Chern number is $C_{dyn}=-\mathit{w}\sin^2 t\Bigr|^{\pi/2}_0=-\mathit{w}$.

\section{Solving the master equation}\label{app:master}
Consider the qubit under the influence of the electromagnetic field, such process can be described by the quantum master equation \cite{Carollo2003,Breuer}

\begin{eqnarray}
	\frac{d\rho}{dt}&=&-i[\mathcal{H}_0(k),\rho]\nonumber\\
	&+&\gamma_0(N+1)(\sigma_-\rho\sigma_+-\frac{1}{2}\left\{ \sigma_+\sigma_-,\rho\right\}) \nonumber\\
	&+&\gamma_0N(\sigma_+\rho\sigma_--\frac{1}{2}\left\{ \sigma_-\sigma_+,\rho\right\})\nonumber
\end{eqnarray}
where the symbols are explained below Eq. (\ref{eq:lindblad}) in the main text.

 The density matrix can be expanded as  $\rho=\frac{1}{2}(I+\vec{a}\cdot\sigma)$ and the master equation [Eq. (\ref{eq:lindblad})] can be written as the following form \cite{Zhang2021}
 \begin{eqnarray}
 	\frac{d}{dt}
 	\begin{pmatrix}
 	a_x\\
 	a_y\\
 	a_z
 	\end{pmatrix}
 	=
 \mathcal{L}(k)
 	\begin{pmatrix}
 	a_x\\
 	a_y\\
 	a_z
 	\end{pmatrix}
 	+
 	\begin{pmatrix}
 	0\\
 	0\\
 	-\gamma_0
 	\end{pmatrix},
 \end{eqnarray}
 where
 \begin{eqnarray}
 \mathcal{L}(k)=	\begin{pmatrix}
 	-\gamma/2& 0 & 2h_y(k) \\
 	0&-\gamma/2&-2h_x(k)\\
 	-2h_y(k)& 2h_x(k)& -\gamma
 \end{pmatrix}.
 \label{eq:eom}
 \end{eqnarray}
To solve Eq. (\ref{eq:eom}), we follow Ref. \cite{Breuer, Zhang2021, Noh2010}. First, we find the stationary solution where the time derivative of the density matrix is zero. The stationary solution is 
\begin{eqnarray}
	a_x^0=\frac{-4h_y\gamma_0}{4
		E_k^2+\gamma^2}\\
	a_y^0=\frac{4h_x\gamma_0}{4E_k^2+\gamma^2}\\
	a_z^0=\frac{-\gamma_0\gamma}{4E_k^2+\gamma^2}.
\label{eq:stationar}
\end{eqnarray}
Second, we define $\vec{a}^c=\vec{a}-\vec{a}^0$ such that $\vec{a}^c$ satisfies the homogeneous differential equation
\begin{eqnarray}
	\frac{d\vec{a}^c}{dt}=\mathcal{L}(k) \vec{a}^c
\end{eqnarray} 
 The left and right eigenvalues of $\mathcal{L}(k)$ are $\lambda_0=\frac{-\gamma}{2}$, $\lambda_{1,2}=\frac{-3\gamma}{4}\mp\frac{i\omega}{4}$, where $\omega=\sqrt{64E_k^2-\gamma^2}$. 
 Thus, the solution to $\vec{a}^c$ is $\vec{a}^c=\sum_{\ell} S_{\ell}|R \ell\rangle e^{\lambda_{\ell} t}$ with $S_{\ell}=\langle L \ell|{a}^c(t=0)\rangle$, where $\langle L \ell|$ and $| R \ell\rangle$ are the left and right eigenvector of the $\ell$th eigenvalue, respectively, $|{a}^c(t=0)\rangle$ denotes the initial column vector $\vec{a}^c$.  The left and right eigenvectors satisfy $\langle L\ell|R\ell'\rangle=\delta_{\ell, \ell'}$.
 	For the quench protocol considered in the main text, the initial state is given by $\vec{a}^c(t=0)=(0,0,1)^T-\vec{a}^0$. At low temperature, one can approximate $\gamma_0=\gamma$ which gives a simple form of the solution
 	\begin{widetext}
\begin{eqnarray}
 	a_x^c(t)&=&h_y F(\gamma,k,t)\\
 a_y^c(t)&=&-h_x  F(\gamma,k,t)\\
 a_z^c(t)&=&e^{-\frac{3\gamma t}{4}} \frac{2 {\omega} \left(2E_k^2+\gamma^2\right) \cos \left(\frac{{\omega} t}{4}\right)-2 \gamma \left(\gamma^2-14 E_k^2\right) \sin \left(\frac{{\omega} t}{4}\right)}{{\omega} \left(4 E_k^2+\gamma^2\right)}, \\
 \text{where  }
 F(\gamma,k,t)&=&4e^{-\frac{3\gamma t}{4}} \frac{(8E_k^2+5\gamma^2)\sin(\frac{\omega t}{4})+\omega\gamma\cos(\frac{\omega t}{4})}
 {\omega(4E_k^2+\gamma^2)}
\end{eqnarray}
\end{widetext}
Thus, the pseudospinors are
given by $\langle \sigma_i\rangle(t) =Tr\frac{1}{2}\left[\rho\sigma_i
\right]=a_i^c(t)+a_i^0$. 
It can be shown that in the limit $\gamma=0$, $F(\gamma,k,t)=
\frac{\sin(2E_k t)}{E_k}$ and $a_z(t)=\cos(2E_k t)$, which agree with the results in Appendix \ref{app:timeavg}.
Moreover, at $t=0$, $F(\gamma,k,t)=\frac{4\gamma}{4E_k^2+\gamma^2}$ and the solution recovers the initial condition $\vec{a}(t=0)=(0,0,1)^T$
 

%

\end{document}